\documentclass[preprintnumbers,prd,nofootinbib,floats,amssymb,floatfix]{revtex4}
\usepackage{amsfonts}
\usepackage{amsmath}
\usepackage{graphicx}
\usepackage{color}
\usepackage{amssymb}
\usepackage{hyperref}
\hypersetup{colorlinks,linkcolor={cyan},citecolor={cyan},urlcolor={magenta}}
\usepackage[normalem]{ulem}

\begin{document}

\title{Symplectic realization of two interacting spin-two fields in three dimensions}

\author{Omar Rodr\'iguez-Tzompantzi}
\email{omar.tz2701@gmail.com}
\affiliation{ Departamento de Ciencias F\'isicas, Universidad Andres Bello, Sazi\'e 2212, Piso 7, Santiago, Chile.}

\begin{abstract}
We constructed a symplectic realization of the dynamic structure of two interacting spin-two fields in three dimensions.  A significant simplification refers to the treatment of constraints: instead of performing a Hamiltonian analysis $\grave{a}\, la$ Dirac, we worked out a method that only uses properties of the pre-symplectic two-form matrix and its corresponding zero-modes to investigate the nature of constraints and the gauge structure of the theory. For instance, we demonstrate that the contraction of the zero-modes with the potential gradient, yields explicit expressions for the whole set of constraints on the dynamics of the theory, including the symmetrization condition and an explicit relationship between the coupling and cosmological constants. This way, we further identify the necessary conditions for the existence of a unique non-linear candidate for a partially massless theory, using only the expression for the interaction parameters of the model. In the case of gauge structure, the transformation laws for the entire set of dynamical variables are more straightforwardly derived from the structure of the remaining zero-modes; in this sense, the zero-modes must be viewed as the generators of the corresponding gauge transformations. Thereafter, we use an appropriate gauge-fixing procedure, the time gauge, to compute both the quantization brackets and the functional measure on the path integral associated with our model. Finally, we confirm that three-dimensional bi-gravity has two physical degrees of freedom per space point. With the above, we provide a new perspective for a better understanding of the dynamical structure of theories of interacting spin-two fields, which does not require the constraints to be catalogued as first- and second-class ones as in the case of Dirac's standard method.
\end{abstract}
\maketitle

\section{Introduction}
Although Einstein's General Theory of Relativity \cite{Einstein} is accepted as being  the only physical theory that describes the geometrical structure of the space-time and the gravitational dynamics of massive bodies. The greatest puzzles in modern cosmology such as Dark Matter \cite{Bertone} and Dark Energy \cite{Frieman}, which attempt to explain the primordial and late time accelerating expansion of the current universe, have been strong motivators for a plethora of alternative gravity theories beyond original Einstein's General Relativity (GR), both at the ultraviolet and infrared scale \cite{Kolb, Durrer, Clifton}. Besides that, the issue concerning the compatibility between renormalizability and the unitarity of gravity at the quantum regime  \cite{Gerard, Stelle}, has also led many theorists to search for extensions of GR and their three-dimensional cousins; because they are, to our knowledge, another framework for addressing some conceptual features of the four-dimensional GR and some fundamental issues of quantum gravity \cite{Witten, Maloney, Kim}. Nonetheless, modifications and/or extensions to the theory of GR are highly restricted according to Lovelock's theorem \cite{Lovelock, Lovelock2}. Along these lines, given the assumption that GR is essentially the correct theory of gravity leading to the propagation of two physical degrees of freedom (DoF) corresponding to a single spin-two field (graviton) whose rest masses may be exactly zero \cite{Dyson, Weinberg, Goldhaber}, a natural extension  to Einstein's gravity would be the addition of new DoF to the massless graviton in a consistent manner.

Massive gravity has been studied extensively over the past years as a straightforward extension of GR as it contains five DoF which correspond to a single massive spin-two field \cite{Kurt,Deffayet}. Furthermore, the study of massive gravity models has been mainly motivated by offering an alternative to the $\Lambda$CDM cosmological standard model \cite{Babichev, Babichev2, Chu} and by solving the problem of finding a consistent quantum theory of gravity \cite{Oda}. This interest was triggered in particular by the discovery  of a non-linear theory of massive gravity by de Rham, Gabadaze, and Tolley (dRGT) \cite{Rham, Rham2, Rham3} which not only propagates the appropriate five DoF for a massive spin-two field, but is also free of ghost instabilities \cite{Hassan0,Hassan,May}, i.e., unphysical DoF appearing generically at the non-linear level and having unbounded negative energy \cite{Ghost}. Motivated by the advances pioneered by dRGT, Hassan and Rosen (H-R) generalize the dRGT model as a \textit{theory of two interacting spin-two fields}, dubbed as bi-gravity, which describes two gravitons interacting with each other, through non-derivative interaction terms \cite{Hassan1}. It was shown in Refs. \cite{Hassan2, Hassan3} that like the dRGT model the H-R theory also has a correct number of ghost-free DoF; namely it contains seven DoF corresponding to one massive (five DoF) and one massless (two DoF) graviton, in contrast to the five DoF for the massive graviton of dRGT theory. A remarkable feature of the theory of two interacting spin-two fields, is that it fills a gap in the list of consistent field theories for massive and massless fields with spin up to two and represents an extended gravitational theory with a rich phenomenology \cite{Laura, HassaN, Deser, Deser2, Deser3, Higuchi,Bellazzini}. Interestingly enough, three-dimensional theories of two interacting spin-two fields have also attracted much attention in the past a few years as a tool to build effective theories, as they describe the dynamics of spin-two gapped collective excitations observed in certain fractional quantum Hall states \cite{Condence, Condence1, Condence2}. This latest application of models of two interacting spin-two fields in condensed matter physics is of particular interest, as it directly links two broad areas in physics; gravity and condensed matter. In this work, we are predominantly interested in the study of a very attractive three-dimensional version of the theories of two interacting spin-two fields, which is known as Zwei-Dreibein gravity \cite{Bergshoeff, Hamid, Banados}.

Strictly speaking, GR viewed from the canonical perspective is a theory fully governed by constraints \cite{Ashtekar, Peldan}, meaning that massive gravity as well as theories of two interacting spin-two fields must possess the necessary physical constraints to explicitly  eliminate the ghost-like unphysical DoF and to describe the dynamical structure of both types of theories. The most standard approach for determining these constraints, is to use the Dirac algorithm for constrained Hamiltonian systems \cite{Dirac, Henneaux, Regge}. In this approach, the primary constraints appear when the canonical momenta are computed. Consistency of the theory requires that the primary constraints remain on the constraint surface during their evolution. These conditions of consistency will lead to secondary constraints, and so on. Thereafter, the whole set of constraints must be classified into first- and second-class ones. In doing so, the number of physical DoF can be explicitly counted, and a generator of the local gauge transformations can be constructed as a suitable combination of the first-class constraints. Finally, the bracket structure (Dirac's brackets) to quantize a gauge system can be obtained once the second-class constraints are removed.

Broadly speaking, the Hamiltonian analysis for these types of theories has turned out to be substantially complicated by the awkward structure of its interaction potential, which is specifically constructed in terms of a real matrix square-root of $\sqrt{g^{-1}f}$ using the elementary symmetric polynomials \cite{Hassan}. It has been nonetheless argued in Refs. \cite{Kurt,Deffayet} that such a square root structure suggests that the first-order formulation of GR, which uses as fundamental variables tetrads and Lorentz connections instead of metrics, could be the appropriate ones for the formulation of interacting spin-two theories. Even for such models based on the first-order formalism, the canonical analysis still remains quite tedious technically since secondary, tertiary and quartic constraints with complicated Poisson brackets algebra, are present, which make the method difficult to apply. To be more precise, it is still difficult to identify the whole set of physical constraint \cite{Kluson, Kluson2, Kluson3, Alexandrov}, which means that the classification and separation of all the constraints into first- and second-class ones can become non-trivial, which can hide the dynamical structure of these theories. What is more, the explicit form of all constraints plays a significant role if one wishes to solve initial-value problems and to explore the partially massless sectors, which could be relevant for cosmological applications, providing further motivation for their computation. As such, it is natural to foresee any straightforward approach that is helpful in finding the explicit form for the whole set of constraints that will completely dictate the dynamics of these kinds of theories.

There exists, however, a symplectic realization that provides straightforward effective tools for the study of the dynamical structure of Lagrangian field theories, which is geometrically well motivated and is based on the symplectic structure of the space-phase \cite{F-J,Neto, Neto2, Montani, Toms,Omar1, Omar2}, and therefore it is different from Dirac's approach. The advantage of this construction is that we do not need to catalogue the constraints as first- and second-class ones. In this setting, all the constraints are treated at the same footing and the ambiguities mentioned above can be alleviated. Moreover, it does not rely on Dirac's conjecture. Still, several essential elements of a physical theory, such as the structure of all physical constraints, the local gauge symmetry and its generators, the quantization brackets structure, the functional measure for determining the quantum transition amplitude, and the correct number of physical DoF, can be systematically addressed by studying only the properties of the pre-symplectic matrix and its corresponding zero-modes. Formally speaking, we can determine the explicit form of all the physical constraints directly from the zero-modes of the corresponding pre-symplectic matrix through the contraction of these zero-modes with the gradient of the symplectic potential. After identifying the whole set of constraints, we can find out the transformation laws for all the set of dynamical variables corresponding to gauge symmetries, encoded in the remaining zero-modes. This means that such zero-modes are indeed the generators of the local gauge symmetry under which all physical quantities are invariant. Furthermore, by introducing an appropriate gauge-fixing procedure, we can obtain the structure of quantization brackets  between the dynamical variable, the functional measure for determining the quantum transition amplitude and the number of physical DoF. In the light of these results, this symplectic method can provides powerful tools to deal with massive gravity and bi-gravity theories, where secondary, tertiary, or higher-order constraints are present, because it is algebraically simpler than the Dirac algorith \cite{Omar1, Omar2}.

Our goal in this work is to give, in a rather systematic manner, such a symplectic treatment to a theory describing two interacting spin-two fields in three dimensions, dubbed as Zwei-Dreibein gravity \cite{Bergshoeff,Hamid,Banados}, which could shed light on the four-dimensional case. The main building blocks of such a theory are two copies of parity-even cosmological Einstein-Cartan theories glued together by a cubic potential involving the two dreibeins. Beginning with the $2+1$ decomposition of the model, we will derive a pre-symplectic two-form matrix that is degenerated. After extracting the corresponding zero-modes of such a singular matrix, we will be able to identify the explicit form for each physical constraint, including the symmetrization constraints, and a unique condition for the partially massless sector of the theory. Having determined the constraints on the dynamics, we will show that the remaining zero-modes are the generators of the local gauge transformations. Thereafter, by choosing an appropriate gauge-fixing procedure, we will obtain the fundamental brackets structure for the dynamic variables and the functional measure for determining the quantum transition amplitude. Finally, we will also confirm that Zwei-Dreibein gravity has two physical degrees of freedom per space point.

The outline of this paper is as follows. In Section \ref{Action}, we will introduce  the action principle corresponding to
Zwei-Dreibein gravity. Later introducing the action principle, we will perform the $2+1$ decomposition of the action in Section \ref{constraints}, in order to identify the dynamical variables that make up the pre-symplectic matrix. We then show that the symplectic framework applied to Zwei-Dreibein gravity easily allows us to obtain the complete set of physical constraints and the partially massless sector of the theory. In Section \ref{Gauge}, we show that the gauge transformations can be computed using the zero-modes of the pre-symplectic two-form matrix. Then we recovered the Diffeomorphisms and Poincar\'e symmetry by mapping the gauge parameters appropriately. In the penultimate Section \ref{Fixing}, we determine the fundamental quantization brackets, the functional measure on the path integral and the physical degrees of freedom associated with our model, by introducing a gauge-fixing procedure. Finally,  Section \ref{Final}, is devoted to our concluding remarks.

\section{The action principle} 
\label{Action}
We consider  the action for three-dimensional bi-gravity in the first-order formalism \cite{Bergshoeff, Hamid, Banados}. The space-time $\mathcal{M}$ is a three-dimensional oriented smooth manifold and the action is simply given by the sum of two three-dimensional Einstein-Cartan actions with independent deibreins and connections, plus an interaction term:
\begin{eqnarray}
\label{action}
S_{\text{Bi-g}}[A,e, w,l]&=& \int_{\mathcal{M}} \left(e^{I}\wedge R[A]_{I}-\frac{\Lambda_{1}}{6}\epsilon^{IJK}e_{I}\wedge e_{J}\wedge e_{K}+l^{I}\wedge F[w]_{I}-\frac{\Lambda_{2}}{6}\epsilon^{IJK}l_{I}\wedge l_{J}\wedge l_{K}\right.\nonumber\\
&& -\left.\frac{ k_{1}}{2}\epsilon^{IJK}e_{I}\wedge e_{J}\wedge l_{K}-\frac{ k_{2}}{2}\epsilon^{IJK}e_{I}\wedge l_{J}\wedge l_{K}\right),
\end{eqnarray}
with  $\Lambda_{1}=-1/l^{2}$ and $\Lambda_{2}=-1/\widetilde{l}^{2}$ the cosmological constants and, $k_{1}$ and $k_{2}$  the coupling. The fundamental  fields of the action (\ref{action}) are: a pair of dreibein one-forms $e^{I}=e_{\mu}^{I}\mathrm{d}x^{\mu}$ and $l^{I}=l_{\mu}^{I}\mathrm{d}x^{\mu}$, and a pair of dualised spin-connection one-forms $A_{I}=\epsilon_{IJK}A_{\mu}{^{JK}}\mathrm{d}x^{\mu}$ and $w_{I}=\epsilon_{IJK}w_{\mu}{^{JK}}\mathrm{d}x^{\mu}$  valued on the adjoint representation of the Lie group $SO(2, 1)$, so that, it admits an invariant totally anti-symmetric tensor $\epsilon^{IJK}$.  Furthermore, $R^{I}$ and $F^{I}$ are the curvatures of the connections $A^{I}$ and $w^{I}$ severally, which explicitly read  $R^{I}[A]= \mathrm{d}A^{I}+(1/2)\epsilon^{IJK}A_{J}\wedge A_{K}$ and $F^{I}[w]= \mathrm{d}w^{I}+(1/2)\epsilon^{IJK}w_{J}\wedge w_{K}$. In particular, we defined  two types of covariant derivative, acting on internal indices, by mean of the  spin-connections $A$ and $w$, respectively
\begin{equation}
D_{\alpha}V^{I}=\partial_{\alpha}V^{I}+\epsilon^{IJK}A_{\alpha J}V_{K},\quad\nabla_{\alpha}V^{I}=\partial_{\alpha}V^{I}+\epsilon^{IJK}w_{\alpha J}V_{K},
\end{equation}
where $\partial$ is a fiducial derivative operator.  In what follows, we denote the components of three-dimensional spacetime $x$ as $x^{\mu}$, $\mu=0,1,2$ and those of space $\mathbf{x}$ as $x^{a}$, $a=1,2$. Whereas, the Latin capital letters $I$ correspond to Lorentz indices, $I=1,2,3$.

An arbitrary variation of the action functional (\ref{action}) gives a bulk term, which defines the classical equations of motion, and a boundary term in the following form:
\begin{eqnarray}
\delta S_{\text{Bi-g}}[\Phi]=\int_{\mathcal{M}}\mathrm{d} x\left(\mathbf{E}_{\Phi}\delta\Phi+d\mathbf{\Theta}\left[\delta\Phi,\Phi\right]\right),
\end{eqnarray}
where $\mathbf{E}_{\Phi}$ represents equations of motion for all dynamical fields collectively denoted by $\Phi$, $\delta\Phi$ are variations of those fields, and $\mathbf{\Theta}$ is the boundary term. Then the field equations have the form:
\begin{eqnarray}
\label{E1}\mathbf{E}_{A}^{\mu I}&=&\varepsilon^{\mu\alpha\beta}\left(R_{\alpha\beta}^{I}-\frac{1}{2}\Lambda_{1}\epsilon^{IJK}e_{\alpha J}e_{\beta K}-\frac{1}{2}k_{2}\epsilon^{IJK}l_{\alpha J}l_{\beta K}-k_{1}\epsilon^{IJK}e_{\alpha J}l_{\beta K}\right)=0,\\
\label{E2}\mathbf{E}_{w}^{\mu I}&=&\varepsilon^{\mu\alpha\beta}\left(F_{\alpha\beta}^{I}-\frac{1}{2}\Lambda_{2}\epsilon^{IJK}l_{\alpha J}l_{\beta K}-\frac{1}{2}k_{1}\epsilon^{IJK}e_{\alpha J}e_{\beta K}-k_{2}\epsilon^{IJK}l_{\alpha J}e_{\beta K}\right)=0,\\
\label{E3}\mathbf{E}_{e}^{\mu I}&=&\varepsilon^{\mu\alpha\beta}D_{\alpha}e_{\beta}^{I}=0,\\
\label{E4}\mathbf{E}_{l}^{\mu I}&=&\varepsilon^{\mu\alpha\beta}\nabla_{\alpha}l_{\beta}^{I}=0,
\end{eqnarray}
while the boundary term look like
\begin{equation}
\mathbf{\Theta}=\int_{\Sigma}\varepsilon^{\mu\alpha\beta}\left(e_{\alpha I}\delta A_{\beta}^{I}+l_{\alpha I}\delta w_{\beta}^{I}\right)\mathrm{d}{x}_{\mu},
\end{equation}
which is understandable as the divergence of a phase space pre-symplectic potential, from which the corresponding symplectic structure is recognized:
\begin{equation}
\Omega=\int_{\Sigma}\varepsilon^{\mu\alpha\beta}\left(\delta e_{\alpha I}\delta A_{\beta}^{I}+\delta l_{\alpha I}\delta w_{\beta}^{I}\right)\mathrm{d}{x}_{\mu}.
\end{equation}
If we choose $\Sigma$ to be a time slice $x^{0}=\text{const}$, then the symplectic structure takes the form
\begin{equation}
\Omega=\int_{\Sigma} \varepsilon^{0ab}\left(\delta e_{a I}\delta A_{b}^{I}+\delta l_{a I}\delta w_{b}^{I}\right)\mathrm{d}\mathbf{x}.
\end{equation}
The corresponding non-vanishing fundamental Poisson brackets between pairs of basic fields read
\begin{equation}
\label{Poisson}
\{e_{a I}(\mathbf{x}), A_{b}^{J}(\mathbf{y})\}=\varepsilon_{0ab}\eta_{I}^{J}\delta^{2}(\mathbf{x}-\mathbf{y})\quad\text{and}\quad\{l_{a I}(\mathbf{x}), w_{b}^{J}(\mathbf{y})\}=\varepsilon_{0ab}\eta_{I}^{J}\delta^{2}(\mathbf{x}-\mathbf{y}).
\end{equation}
The above Poisson brackets should be employed in a Hamiltonian analysis $\grave{a}\, la\, Dirac$, in order to find out the full structure of the constraints in terms of dreibeins and connections. However, in the following lines, we will discuss a new strategy for investigating the constraint structure and the associated gauge symmetry in Zwei-Dreibein gravity.
 \section{The nature of constraints in the symplectic framework} 
 \label{constraints}
To carry out symplectic analysis, we assume not only that the manifold $\mathcal{M}$ is globally hyperbolic such that it may be foliated as $\Sigma\times\Re$, with $\Sigma$ being a Cauchy's surface without boundary ($\partial\Sigma=0$) and $\Re$ an evolution parameter, but also that simultaneous proper 2+1 decompositions exist for the pair of dreibein $(e,l)$,  as shown in Ref. \cite{Mikica}. Starting from the action (\ref{action}) and making all the indices explicit, we can perform the $2+1$ decomposition of our fields and write the action in the following form (we recall that all spatial boundary terms will be neglected because $\Sigma$ has no boundary):
\begin{eqnarray}
S_{\text{Bi-g}}[e,l,A,w] & = & \int_{\Sigma\times\Re} \mathrm{d}t \mathrm{d}\mathbf{x}\varepsilon^{0ab}\left[e_{I0}\left(R_{ab}^{I}-\frac{1}{2}\Lambda_{1}\epsilon^{IJK}e_{aJ}e_{bK}+\frac{1}{2}\epsilon^{IJK}\left(2k_{1}e_{aJ}+k_{2}l_{aJ}\right)l_{bK}\right)+A_{0I}D_{a}e_{b}^{I}+\dot{A}_{aI}e_{b}^{I}\right.\nonumber \\
&&\left.+ l_{I0}\left(F_{ab}^{I}-\frac{1}{2}\Lambda_{2}\epsilon^{IJK}l_{aJ}l_{bK}+\frac{1}{2}\epsilon^{IJK}\left(2k_{2}l_{aJ}+k_{1}e_{aJ}\right)e_{bK}\right)+w_{0I}\nabla_{a}l_{b}^{I}+\dot{w}_{aI}l_{b}^{I}\right],\label{descomposition}
\end{eqnarray}
where we have defined $R_{ab}{^{I}}=\partial_{a}A_{b}{^{I}}+(1/2)\epsilon^{IJK}A_{aJ}A_{bK}$, $D_{a}e_{b}{^{I}}=\partial_{a}e_{b}{^{I}}+\epsilon^{IJK}A_{aJ}e_{bK}$, $F_{ab}^{I}=\partial_{a}w_{b}{^{I}}+(1/2)\epsilon^{IJK}w_{aJ}w_{bK}$ and $\nabla_{a}l_{b}{^{I}}=\partial_{a}l_{b}{^{I}}+\epsilon^{IJK}w_{aJ}l_{bK}$.  To start our analysis, and without loss of generality, we express the action functional (\ref{descomposition}) in the symplectic form as:
\begin{equation}
S_{\text{Bi-g}}[\xi]= \int_{\mathcal{M}}\mathrm{d}x \mathcal{L}=\int_{\Sigma\times\Re} \mathrm{d}t \mathrm{d}\mathbf{x} \left(a_{i}\left[\xi\right]\dot{\xi}^{i}-\mathcal{V}\left[\xi\right]\right),\label{Lag_Sym}
\end{equation}
which is first-order in time derivative $\dot{\xi}^{i}$, where hereafter $\xi^{i}$ stands for the collection of all the dynamical fields of the theory. The functionals $\xi^{i}$ are usually referred as symplectic variables in the symplectic framework \cite{F-J,Neto, Neto2, Montani, Toms,Omar1, Omar2}.  Furthermore, $a_{i}[\xi]$ are the components of  the so-called canonical one-form $a=a_{i}(\xi)\mathrm{d}\xi^{i}$ as defined in Ref. \cite{F-J}, whereas the functional $\mathcal{V}$ represents the symplectic potential density which could also be identified with the canonical Hamiltonian density $\mathcal{H}$. From now on $\mathrm{d}x$ denotes  the product of the differentials of local coordinates $(x^{\mu})=(t,x^{a})$, so that  $\mathrm{d}x=\mathrm{d}t\mathrm{d}\mathbf{x}$ where $\mathrm{d}\mathbf{x}=\mathrm{d}x^{1}\cdots \mathrm{d}x^{a}$ stands for the product of the differentials of the space coordinates. For Zwei-Dreibein gravity, the set of initial symplectic variables and the components of the canonical one-form can be read easily from the first-order Lagrangian density in Eq. (\ref{descomposition}) as follows
\begin{eqnarray}
{\xi}{^{i}}&=& (A{_{0I}},A{_{aI}},w_{0I},w_{aI},e_{0I},e_{aI},l_{0I},l_{aI}), \label{variables1}\\
{a}{_{i}} &=& ( 0,\varepsilon^{0ab}e_{b}^{I},0,\varepsilon^{0ab}l_{b}^{I}, 0,0,0,0 ).\label{form1}
\end{eqnarray}
On the other hand,  the corresponding symplectic potential density reads
\begin{eqnarray}
\mathcal{V}[\xi] & = &- \varepsilon^{0ab}\left[e_{I0}\left(R_{ab}^{I}-\frac{1}{2}\Lambda_{1}\epsilon^{IJK}e_{aJ}e_{bK}+\frac{1}{2}\epsilon^{IJK}\left(2k_{1}e_{aJ}+k_{2}l_{aJ}\right)l_{bK}\right)+A_{0I}D_{a}e_{b}^{I}\right.\nonumber \\
&&\left.+ l_{I0}\left(F_{ab}^{I}-\frac{1}{2}\Lambda_{2}\epsilon^{IJK}l_{aJ}l_{bK}+\frac{1}{2}\epsilon^{IJK}\left(2k_{2}l_{aJ}+k_{1}e_{aJ}\right)e_{bK}\right)+w_{0I}\nabla_{a}l_{b}^{I}\right].\label{Potential}
\end{eqnarray}
In the symplectic picture, the equations of motion deduced from the variational principle applied to Lagrangian (\ref{Lag_Sym}) can be compactly written in a first-order form as \cite{F-J,Neto, Neto2},
\begin{equation}
\int \mathrm{d}\mathbf{x}\left(\mathcal{F}_{ij}(\mathbf{x},\mathbf{y})\dot{\xi}^{j}(\mathbf{x})-\frac{\delta \mathcal{V}(\mathbf{x})}{\delta\xi^{i}(\mathbf{y})}\right)=0,\label{eqmot}
\end{equation}
At this point, it is important to notice that the dynamics of the theory is then characterized by the called pre-symplectic two-form matrix which is defined as a generalized curl of the canonical one-form, i.e.

\begin{equation}
\mathcal{F}_{ij}(\mathbf{x},\mathbf{y})=\frac{\delta a_{j}(\mathbf{x})}{\delta\xi^{i}(\mathbf{y})}-\frac{\delta a_{i}(\mathbf{y})}{\delta\xi^{j}(\mathbf{x})}.\label{symplectic_matrix}
\end{equation}
Clearly, $\mathcal{F}_{ij}$ is an anti-symmetrical matrix that can be either singular or non-singular. According to the symplectic approach, if the matrix $\mathcal{F}_{ij}$ is non-singular, then its inverse can be computed. As a consequence, the set of equations in Eq. (\ref{eqmot}) can immediately be solved for the time evolution of the fields $\xi_{i}$, as follows,
\begin{equation}
\dot{\xi}^{i}(\mathbf{y})=\int \mathrm{d}\mathbf{z}(\mathcal{F}^{ij}(\mathbf{y},\mathbf{z}))^{-1}\int \mathrm{d}\mathbf{x}\frac{\delta \mathcal{V}(\mathbf{x})}{\delta\xi^{j}(\mathbf{z})}.
\end{equation}
In our case, introducing the symplectic variables (\ref{variables1}) and the canonical one-form (\ref{form1}) into the pre-symplectic matrix definition (\ref{symplectic_matrix}), we find that the corresponding pre-symplectic matrix  turns out to be 
\begin{eqnarray}
\label{matrx_sym}\mathcal{F}_{ij}(\mathbf{x},\mathbf{y})=
\varepsilon^{0ab}
  \begin{pmatrix}
 0      &  0   &  0     &  0  &  0 &0&0&0	 	 \\                                                                        
0    &  0      &  0   &   0   &  0&\eta_{IJ}&0&0 	\\
    0      &  0    & 0  & 0	 &  0&0&0&0 \\
0  &   0  &0  & 0 	& 0&0&0&\eta_{IJ} 	 \\
0  &   0  &0  & 0 	& 0&0&0&0 	 \\
0  &   -\eta_{IJ}  &0  & 0 	& 0&0&0&0 	 \\
0  &   0  &0  & 0 	& 0&0&0&0 	 \\
0   &  0  & 0   &  -\eta_{IJ}	&  0&0&0&0 
 \end{pmatrix}
\delta^{2}(\mathbf{x}-\mathbf{y}).\label{f0}
\end{eqnarray}
After constructing the pre-symplectic matrix, we can see clearly that it is manifestly degenerate in the sense that there are more degrees of freedom in the equations of motion (\ref{eqmot}) than physical degrees of freedom in the theory. As a consequence of this, there exist constraints that must remove the unphysical degrees of freedom. In the symplectic framework, the constraints emerge as algebraic relations necessary to maintain the consistency of the equations of motion. Since the matrix (\ref{f0}) is a singular one, it is straightforward to determine that it has the following the zero-modes:
\begin{eqnarray}
v_{1}^{i}&=&\left(v^{A_{0}},0,0,0,0,0,0,0\right),\\
v_{2}^{i}&=&\left(0,0,v^{w_{0}},0,0,0,0,0\right),\\
v_{3}^{i}&=&\left(0,0,0,0,v^{e_{0}},0,0,0\right),\\
v_{4}^{i}&=&\left(0,0,0,0,0,0,v^{l_{0}},0\right),
\end{eqnarray}
where $v^{A{_{0}}}$, $v^{w{_{0}}}$, $v^{e{_{0}}}$, $v^{l{_{0}}}$ are totally arbitrary functions. In terms of the symplectic formalism, the zero-modes satisfy the equation $v_{1,2,3,4}^{i}\mathcal{F}_{ij}=0$, and therefore, the multiplication of each of the zero-modes by the gradient of the symplectic potential (\ref{Potential}) leads to the following conditions:
\begin{eqnarray}
\int \mathrm{d}\mathbf{y}\, v_{1}^{i}(\mathbf{y})\int\ \mathrm{d}\mathbf{x}\frac{\delta\mathcal{V}(\mathbf{x})}{\delta\xi^{i}(\mathbf{y})}  &=&\int \mathrm{d}\mathbf{y}\,v^{A{_{0}}}\varepsilon^{0ab}D_{a}e_{b}^{I}=0,\\\label{p1}
\int \mathrm{d}\mathbf{y}\, v_{2}^{i}(\mathbf{y})\int\ \mathrm{d}\mathbf{x}\frac{\delta\mathcal{V}(\mathbf{x})}{\delta\xi^{i}(\mathbf{y})}  &=&\int \mathrm{d}\mathbf{y}\,v^{w{_{0}}}\varepsilon^{0ab}\nabla_{a}l_{b}^{I}=0,\\\label{p2}
\int \mathrm{d}\mathbf{y}\, v_{3}^{i}(\mathbf{y})\int\ \mathrm{d}\mathbf{x}\frac{\delta\mathcal{V}(\mathbf{x})}{\delta\xi^{i}(\mathbf{y})}  &=&\int \mathrm{d}\mathbf{y}\,v^{e{_{0}}}\varepsilon^{0ab}\left(R_{ab}^{I}-\frac{1}{2}\Lambda_{1}\epsilon^{IJK}e_{aJ}e_{bK}+\frac{1}{2}\epsilon^{IJK}\left(2k_{1}e_{aJ}+k_{2}l_{aJ}\right)l_{bK}\right)=0,\\\label{p3}
\int \mathrm{d}\mathbf{y}\, v_{4}^{i}(\mathbf{y})\int\ \mathrm{d}\mathbf{x}\frac{\delta\mathcal{V}(\mathbf{x})}{\delta\xi^{i}(\mathbf{y})}  &=&\int \mathrm{d}\mathbf{y}\,v^{l{_{0}}}\varepsilon^{0ab}\left(F_{ab}^{I}-\frac{1}{2}\Lambda_{2}\epsilon^{IJK}l_{aJ}l_{bK}+\frac{1}{2}\epsilon^{IJK}\left(2k_{2}l_{aJ}+k_{1}e_{aJ}\right)e_{bK}\right)=0,\label{p4}
\end{eqnarray}
and since $v^{A{_{0}}}$, $v^{w{_{0}}}$, $v^{e{_{0}}}$, $v^{l{_{0}}}$ are arbitrary functions, we obtain the following constraints on the dynamics of the theory
\begin{eqnarray}
\Phi_{1}^{I}  &=&\varepsilon^{0ab}D_{a}e_{b}^{I}=0,\\\label{p1}
\Phi_{2}^{I}&=&\varepsilon^{0ab}\nabla_{a}l_{b}^{I}=0,\\\label{p2}
\Phi_{3}^{I}&=&\varepsilon^{0ab}\left(R_{ab}^{I}-\frac{1}{2}\Lambda_{1}\epsilon^{IJK}e_{aJ}e_{bK}+\frac{1}{2}\epsilon^{IJK}\left(2k_{1}e_{aJ}+k_{2}l_{aJ}\right)l_{bK}\right)=0,\\\label{p3}
\Phi_{4}^{I}&=&\varepsilon^{0ab}\left(F_{ab}^{I}-\frac{1}{2}\Lambda_{2}\epsilon^{IJK}l_{aJ}l_{bK}+\frac{1}{2}\epsilon^{IJK}\left(2k_{2}l_{aJ}+k_{1}e_{aJ}\right)e_{bK}\right)=0.\label{p4}
\end{eqnarray}
At this point, it is worthwhile noting that the constraints $\Phi^{I}_{1,2,3,4}$ have arisen directly from the projection of the equations of motion (\ref{eqmot}). Therefore, they are valid at all times, in particular, $\dot{\Phi}^{I}_{1,2,3,4}=\frac{d}{dt}\Phi^{I}_{1,2,3,4}=0$ hold. In this sense, $\dot{\Phi}^{I}_{1,2,3,4}=0$ are guaranteed to hold and need not be imposed as extra conditions. Contrast this with a Hamiltonian analysis $\grave{a}\, la$ Dirac,  where such conditions must be imposed additionally. Consequently, we shall combinate the equations of motions (\ref{eqmot}) with the fact that $\dot{\Phi}^{I}_{1,2,3,4}=0$ is automatically satisfied to extract hidden constraints,  as described below.

We note now that $\Phi^{I}_{1,2,3,4}$ depend only on the set of symplectic variables $\xi^{i}$, so the condition $\dot{\Phi}^{I}_{1,2,3,4}=0$ can be written as;
\begin{equation}
\frac{\mathrm{d}{\Omega}_{n}[\xi]}{\mathrm{d}t}=\int\mathrm{d}\mathbf{x}\frac{\delta\Omega_{n}(\mathbf{x})}{\delta\xi^{i}(\mathbf{y})}\dot{\xi^{i}}(\mathbf{y})=0\quad\text{with}\quad\Omega_{n}\in\left(\Phi^{I}_{1},\Phi^{I}_{2},\Phi^{I}_{3},\Phi^{I}_{4}\right). \label{consistency}
\end{equation}
Then, the preservation of the constraints $\Omega_{n}$ (\ref{consistency}), together with the equations of motion (\ref{eqmot}) can be rewritten in the following way
\begin{equation}
\int \mathrm{d}\mathbf{x}\left(\mathcal{F}^{(1)}_{mj}(\mathbf{x},\mathbf{y})\dot{\xi}^{j}(\mathbf{x})-\mathcal{Z}_{m}^{(1)}(\mathbf{x},\mathbf{y})\right)=0,\label{eqmot2}
\end{equation}
such that
\begin{equation}
\mathcal{F}_{mj}^{(1)}=\begin{pmatrix}
\mathcal{F}_{ij}\\
\frac{\delta\Omega_{n}}{\delta\xi^{i}}
\end{pmatrix}\quad\text{and}\quad\mathcal{Z}_{m}^{(1)}=\begin{pmatrix}\frac{\delta \mathcal{V}}{\delta\xi^{i}}\\
0\\
0\\
0\\
0
\end{pmatrix}
\quad\text{with}\quad m=i+n.
\end{equation}
Accordingly, the explicit form of the new matrix $\mathcal{F}_{mj}^{(1)}$ is thus
\begin{eqnarray}
\mathcal{F}_{mj}^{(1)}(\mathbf{x},\mathbf{y})&=&{\small{}
\varepsilon^{0ab}
  \begin{pmatrix}
  0&0&0&0&0&0&0&0\\
 0&0      &  0&0   &0&  \eta^{IJ}     &0&  0   \\   
   0&0&0&0&0&0&0&0\\                                                                     
0&0    &  0&0      &  0&0   &  0& \eta^{IJ}   \\
  0&0&0&0&0&0&0&0\\
    0&-\eta^{IJ}      &  0&0&0    & 0  & 0&0	 \\
      0&0&0&0&0&0&0&0\\
0&0  &   0&-\eta^{IJ}  &0  & 0&0&0 	\\
0&\mathbb{E}_{b}^{IJ}  &   0&0&0&-\mathbb{A}_{b}^{\mathbf{y}IJ}&0&0\\
0&0  &   0&\mathbb{L}_{b}^{IJ}  &0&0  &0& -\mathbb{W}_{b}^{\mathbf{y}IJ}\\
0&\mathbb{A}_{b}^{\mathbf{y}IJ} &0&   0  &0&\left(\Lambda_{1}\mathbb{E}_{b}^{IJ}-k_{1}\mathbb{L}_{b}^{IJ}\right)  &0& -\left(k_{1}\mathbb{E}_{b}^{IJ}+k_{2}\mathbb{L}_{b}^{IJ}\right)  	 \\
0&0   &  0&\mathbb{W}_{b}^{\mathbf{y}IJ}  &0& -\left(k_{1}\mathbb{E}_{b}^{IJ}+k_{2}\mathbb{L}_{b}^{IJ}\right)   &0&  \left(\Lambda_{2}\mathbb{L}_{b}^{IJ}-k_{2}\mathbb{E}_{b}^{IJ}\right)	 
 \end{pmatrix}
}\delta^{2}(\mathbf{x}-\mathbf{y}).\label{f}
\end{eqnarray}
Here we have abbreviated $\mathbb{A}_{b}^{\mathbf{y}IJ}=\partial^{\mathbf{y}}_{b}\eta^{IJ}+\epsilon^{IJK}A_{bK}$, $\mathbb{W}_{b}^{\mathbf{y}IJ}=\partial^{\mathbf{y}}_{b}\eta^{IJ}+\epsilon^{IJK}w_{bK}$, $\mathbb{E}_{b}^{IJ}=\epsilon^{IJK}e_{bK}$, $\mathbb{L}_{b}^{IJ}=\epsilon^{IJK}l_{bK}$. We now can easily verify that $\mathcal{F}^{(1)}_{mj}$ is also a singular matrix that has the following linearly independent zero-modes:
\begin{eqnarray}
\label{v1}v_{1}^{(1)m}&=&\left(0,\mathbb{A}_{a}^{\mathbf{y}IJ},0,0,0,\mathbb{E}_{a}^{IJ},0,0,\eta^{IJ},0,0,0\right)\delta^{2}(x-y),\\
\label{v2}v_{2}^{(1)m}&=&\left(0,0,0,\mathbb{W}_{a}^{\mathbf{y}IJ},0,0,0,\mathbb{L}_{a}^{IJ},0,\eta^{IJ},0,0\right)\delta^{2}(x-y),\\
\label{v3}v_{3}^{(1)m}&=&\left(0,k_{1}\mathbb{L}_{a}^{IJ}-\Lambda_{1}\mathbb{E}_{a}^{IJ},0,k_{1}\mathbb{E}_{a}^{IJ}+k_{2}\mathbb{L}_{a}^{IJ},0,\mathbb{A}_{a}^{\mathbf{y}IJ},0,0,0,0,\eta^{IJ},0\right)\delta^{2}(x-y),\\
\label{v4}v_{4}^{(1)m}&=&\left(0,k_{1}\mathbb{E}_{a}^{IJ}+k_{2}\mathbb{L}_{a}^{IJ},0,k_{2}\mathbb{E}_{a}^{IJ}-\Lambda_{2}\mathbb{L}_{a}^{IJ},0,0,0,\mathbb{W}_{a}^{\mathbf{y}IJ},0,0,0,\eta^{IJ}\right)\delta^{2}(x-y).
\end{eqnarray}
Using the symplectic potential (\ref{Potential}), we find that the matrix $\mathcal{Z}_{m}^{(1)}$ has the form
\begin{eqnarray}
\mathcal{Z}_{m}^{(1)}(\mathbf{x},\mathbf{y})=\varepsilon^{0ab}
\begin{pmatrix}
\Phi_{1}^{I}\\
 e_{0J}\mathbb{A}_{b}^{\mathbf{x}IJ}   +A_{0J}\mathbb{E}_{b}^{IJ}\\
 \Phi_{2}^{I}\\
l_{0J}\mathbb{W}_{b}^{\mathbf{x}IJ}+w_{0J}\mathbb{L}_{b}^{IJ}\\
\Phi_{3}^{I}\\
 A_{0J}\mathbb{A}_{b}^{\mathbf{x}IJ} +e_{0J}\left(k_{1}\mathbb{L}_{b}^{IJ}-\Lambda_{1}\mathbb{E}_{b}^{IJ}\right) +l_{0J}\left(k_{1}\mathbb{E}_{b}^{IJ}+k_{2}\mathbb{L}_{b}^{IJ}\right)\\
 \Phi_{4}^{I}\\
    w_{0J}\mathbb{W}_{b}^{\mathbf{x}IJ} +l_{0J}\left(k_{2}\mathbb{E}_{b}^{IJ}-\Lambda_{2}\mathbb{L}_{b}^{IJ}\right) +e_{0J}\left(k_{1}\mathbb{E}_{b}^{IJ}+k_{2}\mathbb{L}_{b}^{IJ}\right)  \\
0   \\
0   \\
0   \\
0   
 \end{pmatrix}
\delta^{2}(\mathbf{x}-\mathbf{y}).\label{f1}
\end{eqnarray}
Multiplying  the zero-modes in Eqs. (\ref{v1})-(\ref{v4}) to the two sides of Eq. (\ref{eqmot2}), we get the following constraint relations (the integration symbols $\int$ are omitted for simplicity): 
\begin{eqnarray}
v_{1}^{(1) m}\mathcal{Z}^{(1)}_{m} &=&-\epsilon^{IJK}\left(A_{0J}\Phi_{1\, K}-e_{0J}\Phi_{3\, K}\right)-\varepsilon^{\mu\alpha\beta}\left(k_{1}e_{\mu}^{I}+k_{2}l_{\mu}^{I}\right)e_{\alpha J}l_{\beta}^{J}=0,\\\label{p5}
v_{2}^{(1) m}\mathcal{Z}^{(1)}_{m}&=&-\epsilon^{IJK}\left(w_{0J}\Phi_{2\,K}-l_{0J}\Phi_{4\, K}\right)-\varepsilon^{\mu\alpha\beta}\left(k_{1}e_{\mu}^{I}+k_{2}l_{\mu}^{I}\right)e_{\alpha J}l_{\beta}^{J}=0,\\\label{p6}
v_{3}^{(1) m}\mathcal{Z}^{(1)}_{m}&=&\epsilon^{IJK}\left(\left(\Lambda_{1}e_{0J}-k_{1}l_{0J}\right)\Phi_{1\, K}-\left(k_{1}e_{0J}+k_{2}l_{0J}\right)\Phi_{2\, K}-A_{0J}\Phi_{3\, K}\right)\nonumber\\
&&+k_{1}\varepsilon^{\mu\alpha\beta}\left(A_{\mu}^{I}-w_{\mu}^{I}\right)e_{\alpha J}l_{\beta}^{J}+\varepsilon^{\mu\alpha\beta}l_{\mu}^{I}\left(k_{1}e_{\alpha J}+k_{2}l_{\alpha J}\right)\left(A_{\beta}^{J}-w_{\beta}^{J}\right)=0,\\\label{p7}
v_{4}^{(1) m}\mathcal{Z}^{(1)}_{m}&=&\epsilon^{IJK}\left(\left(\Lambda_{2}l_{0J}-k_{2}e_{0J}\right)\Phi_{2\, K}-\left(k_{1}e_{0J}+k_{2}l_{0J}\right)\Phi_{1\, K}-w_{0J}\Phi_{4\, K}\right)\nonumber\\
&&+k_{2}\varepsilon^{\mu\alpha\beta}\left(A_{\mu}^{I}-w_{\mu}^{I}\right)e_{\alpha J}l_{\beta}^{J}+\varepsilon^{\mu\alpha\beta}e_{\mu}^{I}\left(k_{1}e_{\alpha J}+k_{2}l_{\alpha J}\right)\left(A_{\beta}^{J}-w_{\beta}^{J}\right)=0.\label{p8}
\end{eqnarray}
Restricting the above relations to the constraints surface, that is $v_{1,2,3,4}^{(1) m}\mathcal{Z}^{(1)}_{m}|_{\Omega_{n}=0}$, we can identify the following set of integrability conditions:
\begin{eqnarray}
\varepsilon^{\mu\alpha\beta}\left(k_{1}e_{\mu}^{I}+k_{2}l_{\mu}^{I}\right)e_{\alpha J}l_{\beta}^{J}&=&0,\\
k_{1}\varepsilon^{\mu\alpha\beta}\left(A_{\mu}^{I}-w_{\mu}^{I}\right)e_{\alpha J}l_{\beta}^{J}+\varepsilon^{\mu\alpha\beta}l_{\mu}^{I}\left(k_{1}e_{\alpha J}+k_{2}l_{\alpha J}\right)\left(A_{\beta}^{J}-w_{\beta}^{J}\right)&=&0,\\
k_{2}\varepsilon^{\mu\alpha\beta}\left(A_{\mu}^{I}-w_{\mu}^{I}\right)e_{\alpha J}l_{\beta}^{J}+\varepsilon^{\mu\alpha\beta}e_{\mu}^{I}\left(k_{1}e_{\alpha J}+k_{2}l_{\alpha J}\right)\left(A_{\beta}^{J}-w_{\beta}^{J}\right)&=&0,
\end{eqnarray}
whose solution is given by
\begin{eqnarray}
\Psi^{\mu}=\varepsilon^{\mu\alpha\beta}e_{\alpha J}l_{\beta}^{J}&=&0,\label{symmetricity}\\
\Upsilon^{\mu}=\varepsilon^{\mu\alpha\beta}\left(k_{1}e_{\alpha J}+k_{2}l_{\alpha J}\right)\left(A_{\beta}^{J}-w_{\beta}^{J}\right)&=&0.\label{symmetricity2}
\end{eqnarray}
The Eq. (\ref{symmetricity}), known as the symmetrization condition, play a crucial role in the relation of the metric and first-order formulations bi-gravity theories \cite{Kurt,Hamid,Banados,May}. It is straightforward to see that the above equations Eqs. (\ref{symmetricity})-(\ref{symmetricity2}) can be split into four equations;
\begin{eqnarray}
\label{one}\Psi&=&\varepsilon^{0ab}e_{aI}l^{I}_{b}=0,\\
\label{two}\Psi^{a}&=&\varepsilon^{0ab}\left(e_{bI}l^{I}_{0}-e_{0I}l_{b}^{I}\right)=0,\label{Multiplier1}\\
\label{three}\Upsilon&=&\varepsilon^{0ab}\left(k_{1}e_{aI}+k_{2}l_{aI}\right)\left(A_{b}^{I}-w_{b}^{I}\right)=0,\\
\Upsilon^{a}&=&\varepsilon^{0ab}\left[\left(k_{1}e_{bI}+k_{2}l_{bI}\right)\left(A_{0}^{I}-w_{0}^{I}\right)-\left(k_{1}e_{0I}+k_{2}l_{0I}\right)\left(A_{b}^{I}-w_{b}^{I}\right)\right]=0.\label{cuatro}
\end{eqnarray}
In this decomposition we can see that the functions (\ref{two}) and (\ref{cuatro}) have mixed the dynamical variables $(e_{aI}, A_{aI}, l_{aI}, w_{aI})$ with the Lagrangian multipliers $(e_{0}^{I}, l_{0}^{I}, A_{0}^{I},w_{0}^{I})$, hence these relations  are not constraints. Whereas the functions (\ref{one}) and (\ref{three})  constitute a set of two new constraints, as they solely involve dynamical variables. In order to be consistent, the new constraints (\ref{one}) and (\ref{three}) must also be preserved under time evolution, that is $\dot{\Psi}=0$ and $\dot{\Upsilon}=0$. Combining Eq. (\ref{eqmot2}) with the preservation in time of  $\Psi$ and $\Upsilon$, we  will obtain a group of new linear equations
\begin{equation}
\int \mathrm{d}\mathbf{x}\left(\mathcal{F}^{(2)}_{sj}(\mathbf{x},\mathbf{y})\dot{\xi}(\mathbf{x})^{j}-\mathcal{Z}_{s}^{(2)}(\mathbf{x},\mathbf{y})\right)=0,\label{eqmot3}
\end{equation}
with
\begin{equation}
\mathcal{F}_{sj}^{(2)}=\begin{pmatrix}
\mathcal{F}^{(1)}_{mj}\\
\frac{\delta\Xi_{l}}{\delta\xi^{j}}
\end{pmatrix}\quad\text{and}\quad\mathcal{Z}_{s}^{(2)}=\begin{pmatrix}\mathcal{Z}_{m}^{(1)}\\
0\\
0
\end{pmatrix}.\label{F2}
\end{equation}
Here $\Xi_{l}\in(\Psi,\Upsilon)$ and $s=m+l$. 

Using $\Psi$ and $\Upsilon$ to obtain $\mathcal{F}_{sj}^{(2)}$, we see that the sub-matrix $\left(\delta\Xi/\delta\xi\right)_{l j}$ of $\mathcal{F}_{sj}^{(2)}$ to be
\begin{eqnarray}
\left(\frac{\delta\Xi}{\delta\xi}\right)_{l j}=
 \varepsilon^{0ab}\begin{pmatrix}
0  &   0  &0  & 0 	& 0&l_{b}^{I}&0&-e_{b}^{I} 	 \\
0   &  - \left(k_{1}e_{b}^{I}+k_{2}l_{b}^{I}\right) & 0   &  \left(k_{1}e_{b}^{I}+k_{2}l_{b}^{I}\right)	&  0&k_{1}\left(A_{b}^{I}-w_{b}^{I}\right)&0&k_{2}\left(A_{b}^{I}-w_{b}^{I}\right)
 \end{pmatrix}\delta^{2}(\mathbf{x}-\mathbf{y}),
\end{eqnarray}
whereas $\mathcal{F}_{mj}^{(1)}$ is defined in Eq. (\ref{f}).

Again, one can easily verify that the matrix $\mathcal{F}_{sj}^{(2)}$ in Eq. (\ref{F2}) is also singular, and so, it has the following linearly independent zero-modes:
\begin{eqnarray}
\label{21}v_{1}^{(2)s}&=&\left(v_{1}^{(1)m},0,0\right),\\
\label{22}v_{2}^{(2)s}&=&\left(v_{2}^{(1)m},0,0\right),\\
\label{23}v_{3}^{(2)s}&=&\left(v_{3}^{(1)m},0,0\right),\\
\label{24}v_{4}^{(2)s}&=&\left(v_{4}^{(1)m},0,0\right),\\
\label{25}v_{5}^{(2)s}&=&\left(0,-k_{1}\left(A_{b}^{I}-w_{b}^{I}\right),0,-k_{2}\left(A_{b}^{I}-w_{b}^{I}\right),0,\left(k_{1}e_{b}^{I}+k_{2}l_{b}^{I}\right),0,-\left(k_{1}e_{b}^{I}+k_{2}l_{b}^{I}\right),0,0,0,0,0,\eta^{IJ}\right)\delta^{2}(\mathbf{x}-\mathbf{y}),\\
\label{26}v_{6}^{(2)s}&=&\left(0,-l_{b}^{I},0,e_{b}^{I},0,0,0,0,0,0,0,0,\eta^{IJ},0\right)\delta^{2}(\mathbf{x}-\mathbf{y}).
\end{eqnarray}
If we multiply the first four zero-mode (\ref{21})-(\ref{24}) by $\mathcal{Z}^{(2)}_{s}$, we will find the same constraint obtained previously, whereas from the zero-mode $v_{5}^{(2)s}$ and $v_{6}^{(2)s}$ we have the following constraint relations  (the integration symbols $\int$ are omitted for simplicity):
\begin{eqnarray}
v_{5}^{(2)s}\mathcal{Z}_{s}^{(2)}&=&\left(k_{1}e_{0I}+k_{2}l_{0I}\right)\left(\Phi_{3}^{I}-\Phi_{4}^{I}\right)-\left(A_{0I}-w_{0I}\right)\left(k_{1}\Phi^{I}_{1}+k_{2}\Phi^{I}_{2}\right)\nonumber\\
&&+\frac{1}{2}\varepsilon^{\alpha\beta\mu}\epsilon^{IJK}\left[\left(k_{1}e_{\alpha I}-k_{2}l_{\alpha I}\right)\left(A_{\beta J}-w_{\beta J}\right)\left(A_{\mu K}-w_{\mu K}\right)\right.\nonumber\\
&&+\left(k_{1}e_{\alpha I}+k_{2}l_{\alpha I}\right)\left(k_{1}e_{\beta J}+k_{2}l_{\beta J}\right)\left(e_{\mu K}-l_{\mu K}\right)\nonumber\\
&&\left.+\left(k_{1}e_{\alpha I}+k_{2}l_{\alpha I}\right)\left(\Lambda_{1}e_{\beta J}e_{\mu K}-\Lambda_{2}l_{\beta J}l_{\mu K}\right)\right],\\
v_{6}^{(2)s}\mathcal{Z}_{s}^{(2)}  &=&l_{0}^{I}\Phi_{1I}-e_{0}^{I}\Phi_{2I}+\varepsilon^{0ab}\epsilon^{IJK}e_{aI}\left(w_{0J}-A_{0J}\right)l_{bK}=0.
\end{eqnarray}
These conditions must hold only on the constraints surface, i.e., when $ \Omega_{n},\Xi_{l}= 0$. Imposing the above condition, we obtain the following  scalar relation between coupling and cosmological constants,
\begin{eqnarray}
\varepsilon^{\alpha\beta\mu}\epsilon^{IJK}&&\left[\left(k_{1}e_{\alpha I}-k_{2}l_{\alpha I}\right)\left(A_{\beta J}-w_{\beta J}\right)\left(A_{\mu K}-w_{\mu K}\right)\right.\nonumber\\
&&\left.+\left(k_{1}e_{\alpha I}+k_{2}l_{\alpha I}\right)\left[\left(k_{1}+\Lambda_{1}\right)e_{\beta J}e_{\mu K}-l_{\beta J}l_{\mu K}\left(k_{2}+\Lambda_{2}\right)-2e_{\beta J}l_{\mu K}\left(k_{1}-k_{2}\right)\right]\right]=0,\label{parameters}
\end{eqnarray}
plus a parameters-free relationship mixing the field $(e_{aI},l_{aI})$ with  $(A_{0}^{I},w_{0}^{I})$, 
\begin{equation}
\varepsilon^{0ab}\epsilon^{IJK}e_{aI}\left(w_{0J}-A_{0J}\right)l_{bK}=0.\label{Multiplier2}
\end{equation}
Since the expression in Eq.(\ref{parameters}) mixes the canonical variables $A_{a}^{I}$, $w_{a}^{I}$, $e_{a}^{I}$ and $l_{a}^{I}$  with the Lagrange multipliers $A_{0}^{I}$, $w_{0}^{I}$, $e_{0}^{I}$ and $l_{0}^{I}$, this condition  becomes in an scalar equation, establishing the most general relationship between the coupling parameters of the theory, rather than acts as a new  constraint on the dynamics of the theory. Hence, there are no further constraints in Zwei-Dreibein gravity and therefore our method to obtain the true set of constraints has finished.

Now, it is very important that, according to our result in Eq. (\ref{parameters}), for the conformal case in which the dreibeins are proportional to each other;
\begin{equation}
k_{1}e_{\alpha}^{I}-k_{2}l_{\alpha}^{I}=0\quad\text{with}\quad k_{1},k_{2}\neq0,\label{anzats}
\end{equation}
we obtain from Eq. (\ref{parameters}) the following condition for the parameter values:
\begin{equation}
12k_{1}^{3}\mathbf{e}\left[c^{2}\left(k_{1}+\Lambda_{1}\right)-\left(k_{2}+\Lambda_{2}\right)+2c\left(k_{2}-k_{1}\right)\right]=0\quad\text{with}\quad c=\frac{k_{2}}{k_{1}},\label{solution}
\end{equation}
where we have used $\varepsilon^{\alpha\beta\mu}e_{\alpha}^{I}e_{\beta}^{J}e_{\mu}^{I}=\mathbf{e}\epsilon^{IJK}$ with $\mathbf{e}=\text{det}|e_{\alpha}^{I}|$. At this stage, we already note that the Eq. (\ref{solution}) (first determined in Ref. \cite{HassaN} in the metric formalism via field equations) immediately suggest the following solution:
\begin{equation}
k_{1}=k_{2}=-\Lambda_{1}=-\Lambda_{2}.\label{anzats1}
\end{equation}
An interesting feature of this solution is that it satisfies the Higuchi bound \cite{HassaN}. Moreover, by substituting (\ref{anzats}) and (\ref{anzats1}) into Eq. (\ref{action}), the bi-gravity action simply reduces to two copies of Einstein-Cartan gravity for $e_{\alpha}^{I}$ with a cosmological constant, explicitly,
\begin{equation}
S[e,A,w]=\int_{\mathcal{M}}e^{I}\wedge\left( R[A]_{I}+ F[w]_{I}+\frac{2}{3}\Lambda_{1}\epsilon^{IJK} e_{J}\wedge e_{K}\right).
\end{equation}
Remarkably, only for these parameter values (\ref{anzats1}), we have that the original theory defined in Eq. (\ref{action}) reduces to a \textit{non-linear partially massless bi-gravity theory}, where the mass value of a very special massive spin-two particle, referred to as \textit{partially massless} is $m^{2}=-\Lambda_{1}$ \cite{HassaN,Deser,Deser2,Deser3, Higuchi}. In this case, the partially-massless spin-two field has $1$ degree of freedom more than those of massless field describing 3D pure gravity and $1$ less that those of massive spin-two field in three dimensions.

According to the symplectic approach, once the whole set of physical constraints $\Gamma_{m}$ has been extracted, one can always rearrange the original Lagrangian density, so that each one of the conditions $\dot{\Gamma}_{m}=0$ directly becomes an equation of motion. This is attained by explicitly incorporating all the constraints obtained into the kinetic part of the initial first-order Lagrangian density, by introducing velocities as Lagrange multipliers in place of the standard ones, and restricting the symplectic potential density to the constraints' surface, in the following form \cite{F-J,Neto,Neto2},
\begin{equation}
\mathcal{L}^{\text{new}}=a_{i}\left[\xi\right]\dot{\xi}^{i}+\dot{\eta}_{m}\Gamma^{m}-\mathcal{V}\left[\xi\right]\arrowvert_{\Gamma_{m}},
\end{equation}
where $\dot{\eta}_{m}$ stand for the Lagrangian multipliers relative to $\Gamma_{m}$.  In the above, $\mathcal{L}^{\text{new}}$ is dynamically equivalent to the original Lagrangian density $\mathcal{L}$ given in Eq. (\ref{Lag_Sym}), since they only differ on the explicit condition that the constraints should not evolve on time, i.e., $\dot{\Gamma}_{m}=0$. Nevertheless, $\mathcal{L}^{\text{new}}$ explicitly contains the necessary information to describe all the dynamics of the theory. In the present case, one can observe that $A^{I}_{0}$, $w^{I}_{0}$, $e^{I}_{0}$ and $l^{I}_{0}$ represent a set of Lagrange multipliers implementing the constraints $\Omega_{n}$, so there is no need to  introduce new Lagrangian multipliers for all the constraints of the theory.  Indeed, by simply performing the redefinition $A^{0}_{I}\rightarrow\dot{\lambda}_{1I}$, $w^{0}_{I}\rightarrow\dot{\lambda}_{2I}$, $e^{0}_{I}\rightarrow\dot{\lambda}_{3I}$ and $l^{0}_{I}\rightarrow\dot{\lambda}_{4I}$, we just need to introduce two new Lagrange multipliers asssociated with $\Psi$ and $\Upsilon$, namely, $\dot{\lambda}_{5}$ and $\dot{\lambda}_{6}$. As a result, the new Lagrangian density may be written as:
\begin{equation}
\mathcal{L}^{\text{new}}=\varepsilon^{0ab}\dot{A}_{aI}e_{b}^{I}+\varepsilon^{0ab}\dot{w}_{aI}l_{b}^{I}-\dot{\lambda}_{1I}\Phi^{I}_{1}-\dot{\lambda}_{2I}\Phi^{I}_{2}-\dot{\lambda}_{3I}\Phi^{I}_{3}-\dot{\lambda}_{4I}\Phi^{I}_{4}-\dot{\lambda}_{5}\Psi-\dot{\lambda}_{6}\Upsilon-\mathcal{V}|_{\Omega_{n},\Xi_{l}}\label{Lag_new}.
\end{equation}

Furthermore, we can note that the symplectic potential density, which can also be identified with the total Hamiltonian density $\mathcal{H}$, turns out to be
\begin{equation}
\mathcal{V}=\mathcal{H}=A_{0I}\Phi_{1}^{I}+w_{0I}\Phi_{2}^{I}+e_{0I}\Phi^{I}_{3}+l_{0I}\Phi^{I}_{4}.\label{Hamiltonian}
\end{equation} 
Note that the symplectic potential drop from the Lagrangian density after being evaluated on the constraint surface, along with $\mathcal{H}|_{\Omega_{n},\Xi_{l}}=0$. This fact shows the general covariance of the theory, and therefore, the dynamics will be governed by the constraints. At this stage, the first-order Lagrangian density (\ref{Lag_new}) explicitly contains all the information to describe the constrained dynamics of the three-dimensional bi-gravity theory.  Now the new set of symplectic variables is identified easily as:
 \begin{eqnarray}
{\xi}{^{\text{(new)}\,i}}&=& (A{_{aI}},w_{aI},e_{aI},l_{aI},\lambda_{1I},\lambda_{2I},\lambda_{3I},\lambda_{4I},\lambda_{5},\lambda_{6}) \label{variables2}.
\end{eqnarray}
 This permits us to identify the new components of the canonical one-form
\begin{eqnarray}
{a}{_{i}}^{\text{(new)}} &=& (\varepsilon^{0ab}e_{b}^{I},\varepsilon^{0ab}l_{b}^{I}, 0,0,-\Phi_{1}^{I},-\Phi_{2}^{I},-\Phi_{3}^{I},-\Phi_{4}^{I},-\Psi,-\Upsilon).\label{form2}
\end{eqnarray}
Making use of the definition of the pre-symplectic two-form matrix in Eq. (\ref{symplectic_matrix}), and after a bit of calculation, we can show that the explicit expression for the corresponding square matrix $\mathcal{F}_{ij}^{\text{(new)}}$ is
\begin{eqnarray}
&&\varepsilon^{0ab}{\small{}
\begin{pmatrix}
 0      &  0   &  \eta^{IJ}     &  0  &  -\mathbb{E}_{b}^{IJ} &0&-\mathbb{A}_{b}^{\mathbf{x}IJ}&0&0&\left(k_{1}e_{b}^{I}+k_{2}l_{b}^{I}\right) \\                                                                        
0    &  0      &  0   &   \eta^{IJ}   &  0&-\mathbb{L}_{b}^{IJ}&0&\mathbb{W}_{b}^{\mathbf{x}IJ}&0&-\left(k_{1}e_{b}^{I}+k_{2}l_{b}^{I}\right)\\
    -\eta^{IJ}      &  0    & 0  & 0	 &  \mathbb{A}_{b}^{\mathbf{x}IJ}&0&-\mathbb{S}_{1b}^{IJ}&\mathbb{K}_{b}^{IJ}&-l_{b}^{I}&-k_{1}\left(A_{b}^{I}-w_{b}^{I}\right) \\
0  &   -\eta^{IJ}  &0  & 0 	& 0&\mathbb{W}_{b}^{\mathbf{x}IJ}&\mathbb{K}_{b}^{IJ}&-\mathbb{S}_{2b}^{IJ}&e_{b}^{I}&-k_{2}\left(A_{b}^{I}-w_{b}^{I}\right)\\
\mathbb{E}_{b}^{IJ}  &   0  &-\mathbb{A}_{b}^{\mathbf{y}IJ} & 0 & 0&0&0&0&0&0\\
0  &  \mathbb{L}_{b}^{IJ}  &0  & -\mathbb{W}_{b}^{\mathbf{y}IJ}& 0&0&0&0&0&0\\
\mathbb{A}_{b}^{\mathbf{y}IJ} &   0  &\mathbb{S}_{1b}^{IJ} & -\mathbb{K}_{b}^{IJ}& 0&0&0&0&0&0 	 \\
0   &  \mathbb{W}_{b}^{\mathbf{y}IJ}  & -\mathbb{K}_{b}^{IJ} &  \mathbb{S}_{2b}^{IJ}&  0&0&0&0&0&0\\
0   &  0  & l_{b}^{I} &  -e_{b}^{I}&  0&0&0&0&0&0\\
\left(k_{1}e_{b}^{I}+k_{2}l_{b}^{I}\right)  &  -\left(k_{1}e_{b}^{I}+k_{2}l_{b}^{I}\right)   & k_{1}\left(A_{b}^{I}-w_{b}^{I}\right) &  k_{2}\left(A_{b}^{I}-w_{b}^{I}\right)&  0&0&0&0&0&0
\end{pmatrix}}
 \nonumber\\
&&\times \delta^{2}(\mathbf{x}-\mathbf{y}).\label{ff}
\end{eqnarray}
Here we have abbreviated $\mathbb{S}_{1 b}^{IJ}=\Lambda_{1}\mathbb{E}_{b}^{IJ}-k_{1}\mathbb{L}_{b}^{IJ}$, $\mathbb{S}_{2 b}^{IJ}=\Lambda_{2}\mathbb{L}_{b}^{IJ}-k_{2}\mathbb{E}_{b}^{IJ}$ and $\mathbb{K}_{b}^{IJ}= k_{1}\mathbb{E}_{b}^{IJ}+k_{2}\mathbb{L}_{b}^{IJ}$. Notwithstanding the above positive results, it is worth noting that the corresponding pre-symplectic matrix $\mathcal{F}_{ij}^{\text{(new)}}$ remains singular, however, we have shown that no further constraints are generated by the zero-modes. Thus, the above observation implies that there might be gauge degrees of freedom in the theory that must be fixed through gauge conditions in order to obviate the singularity. In this way, the quantization-bracket structure can be determined and the procedure can be achieved in terms of the physical degrees of freedom.

\section{Gauge transformations and its generators}
\label{Gauge}
We thus proceed towards the discussion of the gauge symmetry in the symplectic framework. It is worth noting that, 
the degeneracy of the pre-symplectic matrix (\ref{ff}) and the fact that its remaining zero-modes are orthogonal to the gradient of the potential that means that the remaining zero-modes generate degenerate gauge directions inside the symplectic potential (\ref{Hamiltonian}).  As a consequence, such zero-modes must be identified as the generators of the corresponding gauge symmetry `$\delta_{G}$', that is, the components of the zero-modes give the transformation properties related to the underlying (gauge) symmetry that leaves the action invariant \cite{Neto, Montani}. The local infinitesimal transformations on the symplectic variables generated by $v^{(\text{new})i}_{A}$ can be expressed as:
\begin{equation}
\delta_{G}\xi^{i}=\int \mathrm{d}\mathbf{x}\, v_{A}^{(\text{new})i}\eta^{A}, \label{genarador_gauge}
\end{equation}
where $v^{(\text{new})i}_{A}$ are the independent zero-modes of the singular symplectic matrix $\mathcal{F}_{ij}^{\text{(new)}}$ and $\eta^{A}$ are the gauge parameters. For the singular pre-symplectic matrix (\ref{ff}),  these zero-modes turn out to be:
\begin{eqnarray}
v_{1}^{(\text{new})i}&=&\left(\mathbb{A}_{a}^{\mathbf{y}IJ},0,\mathbb{E}_{a}^{IJ},0,\eta^{IJ},0,0,0,0,0\right)\delta^{2}(x-y),\\
v_{2}^{(\text{new})i}&=&\left(0,\mathbb{W}_{a}^{\mathbf{y}IJ},0,\mathbb{L}_{a}^{IJ},0,\eta^{IJ},0,0,0,0\right)\delta^{2}(x-y),\\
v_{3}^{(\text{new})i}&=&\left(-\left(\Lambda_{1}\mathbb{E}_{a}^{IJ}-k_{1}\mathbb{L}_{a}^{IJ}\right),\left(k_{1}\mathbb{E}_{a}^{IJ}+k_{2}\mathbb{L}_{a}^{IJ}\right),\mathbb{A}_{a}^{\mathbf{y}IJ},0,0,0,\eta^{IJ},0,0,0\right)\delta^{2}(x-y),\\
v_{4}^{(\text{new})i}&=&\left(\left(k_{1}\mathbb{E}_{a}^{IJ}+k_{2}\mathbb{L}_{a}^{IJ}\right),-\left(\Lambda_{2}\mathbb{L}_{a}^{IJ}-k_{2}\mathbb{E}_{a}^{IJ}\right),0,\mathbb{W}_{a}^{\mathbf{y}IJ},0,0,0,\eta^{IJ},0,0\right)\delta^{2}(x-y),
\end{eqnarray}
these zero-modes turn out to be orthogonal to the gradient of the symplectic potential and at the same time generate local displacements on the isopotential surface. In this way, one can infer from Eq. (\ref{genarador_gauge}), that the infinitesimal gauge transformations that leave the original Lagrangian invariant take the form
\begin{eqnarray}
\label{gauge1}\delta_{G}A_{\alpha}^{I}&=&-D_{\alpha}\eta^{I}-\epsilon^{IJK}\varsigma_{J}\left(\Lambda_{1}e_{\alpha K}-k_{1}l_{\alpha K}\right)+\epsilon^{IJK}\sigma_{J}\left(k_{1}e_{\alpha K}+k_{2}l_{\alpha K}\right),\\
\label{gauge2}\delta_{G}w_{\alpha}^{I}&=&-\nabla_{\alpha}\varrho^{I}-\epsilon^{IJK}\sigma_{J}\left(\Lambda_{2}l_{\alpha K}-k_{2}e_{\alpha K}\right)+\epsilon^{IJK}\varsigma_{J}\left(k_{1}e_{\alpha K}+k_{2}l_{\alpha K}\right),\\
\label{gauge3}\delta_{G}e_{\alpha}^{I}&=&-D_{\alpha}\varsigma^{I}+\epsilon^{IJK}\eta_{J}e_{\alpha K},\\
\label{gauge4}\delta_{G}l_{\alpha}^{I}&=&-\nabla_{\alpha}\sigma^{I}+\epsilon^{IJK}\varrho_{J}l_{\alpha K}.
\end{eqnarray}
where $\eta^{I}$, $\varsigma^{I}$, $\sigma^{I}$ and $\varrho^{I}$ are the arbitrary gauge transformation parameter. It is worth remarking that (\ref{gauge1}), (\ref{gauge2}), (\ref{gauge3}) and (\ref{gauge4}) correspond to the fundamental gauge symmetry of the theory, though the diffeomorphisms symmetry `$\delta_{\textit{Diff}}$' have not been found yet. However, it is well-known that an appropriate choice of the gauge parameters does generate the diffeomorphism on-shell. Let us redefine the gauge parameters:
\begin{equation}
\eta^{I}=-A_{\beta}^{I}\zeta^{\beta},\,\varrho^{I}=-w_{\beta}^{I}\zeta^{\beta},\, \varsigma^{I}=-e_{\beta}^{I}\zeta^{\beta},\,\sigma^{I}=-l_{\beta}^{I}\zeta^{\beta},
\end{equation}
with $\zeta^{\beta}$ an arbitrary three-vector. Consequently, from the fundamental gauge symmetries (\ref{gauge1})-(\ref{gauge4}), we obtain
\begin{eqnarray}
\delta_{\textit{Diff}}A_{\alpha}^{I}&=&\mathfrak{L}_{\zeta}A_{\alpha}^{I}+\varepsilon_{\alpha\beta\mu}\zeta^{\beta}\mathbf{E}^{\mu I}_{A},\\
\delta_{\textit{Diff}}w_{\alpha}^{I}&=&\mathfrak{L}_{\zeta}w_{\alpha}^{I}+\varepsilon_{\alpha\beta\mu}\zeta^{\beta}\mathbf{E}^{\mu I}_{w},\\
\delta_{\textit{Diff}}e_{\alpha}^{I}&=&\mathfrak{L}_{\zeta}e_{\alpha}^{I}+\varepsilon_{\alpha\beta\mu}\zeta^{\beta}\mathbf{E}^{\mu I}_{e},\\
\delta_{\textit{Diff}}l_{\alpha}^{I}&=&\mathfrak{L}_{\zeta}l_{\alpha}^{I}+\varepsilon_{\alpha\beta\mu}\zeta^{\beta}\mathbf{E}^{\mu I}_{l}.
\end{eqnarray}
which are precisely diffeomorphisms on-shell. Besides, diffeomorphism invariant theories have the Poincar\'e gauge transformations `$\delta_{\textit{PGT}}$', i.e. local Lorentz rotations and translations, as off-shell symmetries by construction \cite{Utiyama, Nester}. Thus, to recover the Poincar\'e symmetry we need to map the arbitrary gauge parameters of the fundamental gauge symmetry $\delta_{G}$ into those of the Poincar\'e symmetry. This is achieved by a mapping of the gauge parameters \cite{Blagojevic, Ortin, Carlip}, e.g.:
\begin{equation}
\eta^{I}=A_{\beta}^{I}\theta^{\beta}+\overline{\omega}^{I},\,\varrho^{I}=w_{\beta}^{I}\theta^{\beta}+\overline{\omega}^{I},\, \varsigma^{I}=e_{\beta}^{I}\theta^{\beta},\,\sigma^{I}=l_{\beta}^{I}\theta^{\beta}
\end{equation}
such that $\theta^{\mu}$ and $\overline{\omega}^{I}$ are related to local coordinate translations and local Lorentz rotations, respectively, which together constitute the 6 independent gauge parameters of Poincar\'e symmetries in 3D. By using this map, the gauge symmetries reproduce the Poincar\'e symmetries modulo terms proportional to the equations of motion:
\begin{eqnarray}
\delta_{\textit{PGT}}A_{\alpha}^{I}&=&-\partial_{\alpha}\overline{\omega}-\epsilon^{IJK}A_{\alpha J}\overline{\omega}^{I}-A_{\beta}^{I}\partial_{\alpha}\theta^{\beta}-\theta^{\beta}\partial_{\beta}A_{\alpha}^{I}-\varepsilon_{\alpha\beta\mu}\theta^{\beta}\mathbf{E}^{\mu I}_{A},\\
\delta_{\textit{PGT}}w_{\alpha}^{I}&=&-\partial_{\alpha}\overline{\omega}-\epsilon^{IJK}w_{\alpha J}\overline{\omega}^{I}-w_{\beta}^{I}\partial_{\alpha}\theta^{\beta}-\theta^{\beta}\partial_{\beta}w_{\alpha}^{I}-\varepsilon_{\alpha\beta\mu}\theta^{\beta}\mathbf{E}^{\mu I}_{w},\\
\delta_{\textit{PGT}}e_{\alpha}^{I}&=&-e_{\beta}^{I}\partial_{\alpha}\theta^{\beta}-\theta^{\beta}\partial_{\beta}e_{\alpha}^{I}-\epsilon^{IJK}e_{\alpha J}\overline{\omega}^{I}-\varepsilon_{\alpha\beta\mu}\theta^{\beta}\mathbf{E}^{\mu I}_{e},\\
\delta_{\textit{PGT}}l_{\alpha}^{I}&=&-l_{\beta}^{I}\partial_{\alpha}\theta^{\beta}-\theta^{\beta}\partial_{\beta}l_{\alpha}^{I}-\epsilon^{IJK}l_{\alpha J}\overline{\omega}^{I}-\varepsilon_{\alpha\beta\mu}\theta^{\beta}\mathbf{E}^{\mu I}_{l}.
\end{eqnarray}
We thus conclude that the Poincar\'e symmetry $\delta_{PGT}$, as well as the diffeomorphisms $\delta_{Diff}$, are not independent symmetries: they are contained in the fundamental gauge symmetry $\delta_{G}$ as on-shell, that is, only when the equations of motion are imposed. Besides, the generators of such gauge transformations are represented in terms of the zero-modes, thereby making evident that the zero-modes of the pre-symplectic two-form encode all the information about the gauge structure of this theory.

\section{Fixing gauge}
\label{Fixing}
As was already mentioned in Sec. \ref{constraints}, in theories with gauge symmetry, the pre-symplectic matrix obtained at the end of the procedure is still singular. Nevertheless, to obtain a non-singular symplectic matrix and to determine the quantization-bracket structure between the dynamical fields, we must impose a gauge-fixing procedure, that is, new gauge constraints. In this case, we now partially fix the gauge, the most natural manner is to choose the time-gauge, namely, $A_{0}^{I}=0$ and $e_{0}^{I}=0$. But since the variables $A^{I}_{0}$ and $e^{I}_{0}$ have been redefined as $A^{0}_{I}\rightarrow\dot{\lambda}_{1I}$ and  $e^{0}_{I}\rightarrow\dot{\lambda}_{3I}$, we can replace the above gauge fixing conditions by $\lambda_{1I}=constant$ and $\lambda_{3I}=constant$, respectively. As before, we can also introduce new Lagrange multipliers relative to the gauge fixing conditions, namely, $\dot{\lambda}_{7 I}$ and $\dot{\lambda}_{8I}$.  Hence, up to a  total time derivative term which does not affect the equations of motion, the resulting Lagrangian density after gauge fixing has the following structure,
\begin{equation}
\mathcal{L}^{\text{final}}=\varepsilon^{0ab}\dot{A}_{aI}e_{b}^{I}+\varepsilon^{0ab}\dot{w}_{aI}l_{b}^{I}-\dot{\lambda}_{1I}\left(\Phi^{I}_{1}-\lambda_{7I}\right)-\dot{\lambda}_{2I}\Phi^{I}_{2}-\dot{\lambda}_{3I}\left(\Phi^{I}_{3}-\lambda_{8I}\right)-\dot{\lambda}_{4I}\Phi^{I}_{4}-\dot{\lambda}_{5}\Psi-\dot{\lambda}_{6}\Upsilon.\label{Lag_final}
\end{equation}
Clearly, from the Lagrangian density (\ref{Lag_final}), the final set of symplectic variables  is taken as 
\begin{eqnarray}
{\xi}{^{i\, \text{(final)} }}&=& (A{_{aI}},w_{aI},e_{aI},l_{aI},\lambda_{1I},\lambda_{2I},\lambda_{3I},\lambda_{4I},\lambda_{5},\lambda_{6},\lambda_{7I},\lambda_{8I}), \label{variables3}\\
{a}{_{i}}^{\text{(final)}} &=& (\varepsilon^{0ab}e_{b}^{I},\varepsilon^{0ab}l_{b}^{I}, 0,0,-\Phi_{1}^{I}+\lambda_{7I},-\Phi_{2}^{I},-\Phi_{3}^{I}+\lambda_{8I},-\Phi_{4}^{I},-\Psi,-\Upsilon,0,0).\label{form3}
\end{eqnarray}
From the above set of the symplectic variables, we finally obtain the pre-symplectic matrix defined in Eq. (\ref{symplectic_matrix}), given by a block matrix of the form
\begin{equation}
\mathcal{F}_{ij}^{\text{(final)}}=
\label{Ff}\begin{pmatrix}
 \mathbf{A}  &  \mathbf{B}  \\                                                                        
\mathbf{C}   &  \mathbf{D} 
\end{pmatrix}\quad\text{with}\quad\mathbf{C}=-\mathbf{B} ^{T}.
\end{equation}
Here the explicit form of each sub matrix $\mathbf{A}$, $\mathbf{B}$ and $\mathbf{D}$ in Eq. (\ref{Ff}) turns out to be
\begin{eqnarray}
\mathbf{A}=\varepsilon^{0ab}
\begin{pmatrix}
 0      &  0   &  \eta^{IJ}     &  0    \\                                                                        
0    &  0      &  0   &   \eta^{IJ}   \\
    -\eta^{IJ}      &  0    & 0  & 0 \\
0  &   -\eta^{IJ}  &0  & 0 	
\end{pmatrix}\delta^{2}(\mathbf{x}-\mathbf{y}),
\end{eqnarray}

\begin{eqnarray}
\mathbf{B}=\varepsilon^{0ab}
\begin{pmatrix}
  -\mathbb{E}_{b}^{IJ} &0&-\mathbb{A}_{b}^{\mathbf{x}IJ}&0&0&-\frac{1}{2}\epsilon^{IJK}\mathbb{K}_{bJK} &0&0\\                                                                        
  0&-\mathbb{L}_{b}^{IJ}&0&-\mathbb{W}_{b}^{\mathbf{x}IJ}&0&\frac{1}{2}\epsilon^{IJK}\mathbb{K}_{bJK}&0&0\\
  \mathbb{A}_{b}^{\mathbf{x}IJ}&0&-\mathbb{S}_{1b}^{IJ}&\mathbb{K}_{b}^{IJ}&-l_{b}^{I}&-k_{1}\left(A_{b}^{I}-w_{b}^{I}\right)&0&0 \\
 0&\mathbb{W}_{b}^{\mathbf{x}IJ}&\mathbb{K}_{b}^{IJ}&-\mathbb{S}_{2b}^{IJ}&e_{b}^{I}&-k_{2}\left(A_{b}^{I}-w_{b}^{I}\right)&0&0
 \end{pmatrix} \delta^{2}(\mathbf{x}-\mathbf{y}),
\end{eqnarray}
\begin{eqnarray}
\mathbf{D}=\varepsilon^{0ab}
\begin{pmatrix}
 0&0&0&0&0&0&-\frac{1}{2}\varepsilon_{0ab}\eta^{IJ}&0\\
 0&0&0&0&0&0&0&0\\
0&0&0&0&0&0&-\frac{1}{2}\varepsilon_{0ab}\eta^{IJ} 	 \\
  0&0&0&0&0&0&0&0\\
  0&0&0&0&0&0&0&0\\
0&0&0&0&0&0&0&0\\
  \frac{1}{2}\varepsilon_{0ab}\eta^{IJ}&0&0&0&0&0&0&0\\
0&0&\frac{1}{2}\varepsilon_{0ab}\eta^{IJ}&0&0&0&0&0
 \end{pmatrix}
 \delta^{2}(\mathbf{x}-\mathbf{y}).\label{f1}
\end{eqnarray}
Here, we can easily see that $\mathbf{A}$ is invertible. 

Using the standard identity for any matrix
\begin{eqnarray}
\label{det}
\left(
  \begin{array}{cc}
  \mathbf{A}	   & \mathbf{B}	  	  	\\
 \mathbf{C}  	 &	  \mathbf{D}     	\\
 \end{array}
\right)=\left(
  \begin{array}{cc}
  \mathbf{A}	   & {\bf 0}	  	  	\\
 \mathbf{C}  	 &	  {\bf 1}     	\\
 \end{array}
\right)\left(
  \begin{array}{cc}
  {\bf 1}	   & \mathbf{A}^{-1}\mathbf{B}	  	  	\\
 {\bf 0} 	 &	  \mathbf{D}-\mathbf{C}\mathbf{A}^{-1}\mathbf{B}     	\\
 \end{array}
\right),
\end{eqnarray}
where $\mathbf{A}$ and $\mathbf{D}$ are square matrices, but $\mathbf{B}$ and $\mathbf{C}$ need not be square. We can see that
\begin{equation}
\text{det}\, \left(
  \begin{array}{cc}
  \mathbf{A}	   & \mathbf{B}	  	  	\\
 \mathbf{C}  	 &	  \mathbf{D}     	\\
 \end{array}
\right)=\left(\text{det}\, \mathbf{A}\right)\left(\text{det}\,\left[\mathbf{D}-\mathbf{C}\mathbf{A}^{-1}\mathbf{B}\right]\right).\label{compute_det}
\end{equation}
Hence, making use of (\ref{compute_det}), and after some algebra, it is possible to show that the determinant of $\mathcal{F}^{\text{(final)}}_{ij}$ is
\begin{equation}
\text{det}\left|\mathcal{F}_{ij}^{\text{(final)}}\right|^{1/2}=\left(k_{1}\Delta+2k_{2}\right)\left|2\Delta\left(k_{1}\Delta+2k_{2}\right)+2\left(k_{2}\Delta-2\Lambda_{2}\right)+\left(2w^{2}-3\partial^{2}\right)\right|\neq0,
\end{equation}
here we have defined $\Delta=e_{c}^{J}l^{c}_{J}$.  This result leads us to conclude that $\mathcal{F}^{\text{(final)}}_{ij}$ is not singular, and therefore the inverse of this matrix exists: it is dubbed as the symplectic two-form matrix.  Now, it is interesting to note that according to Toms approach \cite{Toms}, the functional measure on the path integral associated with our model, in the time gauge, is
\begin{equation}
d\mu=\left(\prod_{i} \left[D \xi^{\text{(final)}}_{i}\right]\right)\left(\text{det}\,\mathcal{F}^{\text{(final)}}_{ij}(\mathbf{x},\mathbf{y})\right)^{1/2}.
\end{equation}

On the other hand, since the matrix (\ref{Ff}) is of block form, its formal inversion is of the form
\begin{equation}
\label{symplectic}
\left(\mathcal{F}_{ij}^{\text{(final)}}\right)^{-1}=
\begin{pmatrix}
 \mathbf{A}^{-1}+\mathbf{A}^{-1}\mathbf{B}\left(\mathbf{D}-\mathbf{C}\mathbf{A}^{-1}\mathbf{B}\right)^{-1}\mathbf{C}\mathbf{A}^{-1} & \quad -\mathbf{A}^{-1}\mathbf{B}\left(\mathbf{D}-\mathbf{C}\mathbf{A}^{-1}\mathbf{B}\right)^{-1}  \\                                                                        
-\left(\mathbf{D}-\mathbf{C}\mathbf{A}^{-1}\mathbf{B}\right)^{-1}\mathbf{C}\mathbf{A}^{-1}   &\quad \left(\mathbf{D}-\mathbf{C}\mathbf{A}^{-1}\mathbf{B}\right)^{-1}
\end{pmatrix}.
\end{equation}
In this setup, the form of the above matrix $\left(\mathcal{F}_{ij}^{\text{(final)}}\right)^{-1}$ defines the brackets $\{\bullet, \bullet\}$, dubbed as the Faddev-Jackiw fundamental brackets, between any two elements of the symplectic variables set $\xi_{i}^{\text{(final)}}$ over the phase space through 
\begin{equation}
\{\xi_{i}^{\text{(final)}}(\mathbf{x}),\xi_{j}^{\text{(final)}}(\mathbf{y})\}=\left(\mathcal{F}_{ij}^{\text{(final)}}\left(\mathbf{x},\mathbf{y}\right)\right)^{-1}.
\end{equation}
After some algebraic manipulations, we have
\begin{eqnarray}
\{A_{a}^{I},A_{b}^{J}\}&=&\{A_{a}^{I},w_{b}^{J}\}=\{A_{a}^{I},e_{b}^{J}\}=\{A_{a}^{I},l_{b}^{J}\}=0,\\
\{w_{a}^{I},w_{b}^{J}\}&=&\{w_{a}^{I},e_{b}^{J}\}=\{w_{a}^{I},l_{b}^{J}\}=0,\\
\{e_{a}^{I},e_{b}^{J}\}&=&\{e_{a}^{I},l_{b}^{J}\}=0,\\
\{l_{a}^{I},l_{b}^{J}\}&=&0.
\end{eqnarray}
Then the above brackets all turn out to be either zero unlike those derived in Eq. (\ref{Poisson}). Now, we need a bracket for observables on $\Sigma$. Such a bracket should agree with the commutator in the classical limit. In this regard, for any two observable $\mathcal{O}_{1}$, $\mathcal{O}_{2}$ defined on the phase space which owns itself a symplectic structure as $\left\{\xi_{i}^{\text{(final)}},\xi_{j}^{\text{(final)}}\right\}$, we can define the following relation:
\begin{equation}
\{\mathcal{O}_{1}(\xi),\mathcal{O}_{2}(\xi)\}=\sum_{i,j}\int d^{3}\mathbf{r}\frac{\delta \mathcal{O}_{1}(\xi)}{\delta\xi_{i}(\mathbf{r})}\left(\mathcal{F}_{ij}^{\text{(final)}}\right)^{-1}\frac{\delta \mathcal{O}_{2}(\xi)}{\delta\xi_{j}(\mathbf{r})}.
\end{equation}
This can be taken as the definition of the Poisson bracket. The canonical quantization can be fully made at tree level by the replacement of classical observables and Poisson brackets by the quantum operators commutators, respectively, according to:
\begin{equation}
\left\{\mathcal{O}_{1}(\mathbf{x}),\mathcal{O}_{2}(\mathbf{y})\right\}\longrightarrow\frac{1}{i\hbar}\left[ \hat{\mathcal{O}}_{1}(\mathbf{x}),\hat{\mathcal{O}}_{2}(\mathbf{y})\right],\,\hat{\mathcal{O}}|\psi\rangle=0,
\end{equation}
where $\hat{\mathcal{O}}$ is any operator associated with an observable (or constraint) and $|\psi\rangle$ is any quantum state.

Finally, the number of propagating degrees of freedom may be calculated in the phase space from the relation
\begin{equation}
N=\frac{1}{2}\left(N_{1}-N_{2}-N_{3}\right)
\end{equation}
where $N_{1}$ is the number of field components in $\xi^{i}=\left(A_{0}^{I},A_{a}^{I},w_{0}^{I},w_{a}^{I},e_{0}^{I},e_{a}^{I},l_{0}^{I},l_{a}^{I}\right)$, $N_{2}$ is the number of fields eliminated $\left(A_{0}^{I},w_{0}^{I},e_{0}^{I},l_{0}^{I}\right)$, and $N_{3}$ is the number constraints including gauge fixing conditions $\left(\Phi_{1}^{I},\Phi_{2}^{I},\Phi_{3}^{I},\Phi_{4}^{I},\Psi,\Upsilon,A_{0}^{I}=0,e_{0}^{I}=0\right)$. Hence, it is concluded that contrary to standard 3D pure gravity, bi-gravity in three dimensions has $1/2\left(36-12-20\right)=2$ physical degrees of freedom; $2$ degrees of freedom for a massive spin-two field and $0$ for a massless spin-2 field in three dimensions, which is the desired result in agreement with \cite{Banados}.
\section{Final remarks}
\label{Final}
In this paper we have discussed the construction of a symplectic realization for a theory describing two interacting spin-two fields in three dimensions, called Zwei-Dreibein gravity. The construction was done within the symplectic framework developed originally by Faddeev and Jackiw \cite{F-J,Omar1,Omar2}. A central point in our discussion is that the analysis be focused on the properties of the pre-symplectic two-form matrix and its corresponding zero-modes which are associated with the constrained dynamics of the theory. The remarkable feature of such a construction is that it does not need to classify the constraints into first- and second-class ones as in the case of the standard Hamiltonian procedures. For instance, using only the zero-modes of the corresponding pre-symplectic matrix, we have explained how to extract, systematically and consistently, the structure of all the physical constraints on the dynamics of the theory. This has led us to infer the necessary conditions under which a candidate for a partially massless theory at the non-linear level exists. Furthermore, upon using the remaining zero-modes, we explicitly found out  the full gauge transformations for all the fundamental variables, while showing that the remaining zero-modes are indeed the generators of the local gauge symmetry under which all physical quantities are invariant. In particular, we successfully recovered the Diffeomorphisms and Poincar\'e symmetry by mapping the gauge parameters appropriately. This leads to a significant reduction of labor compared to the framework of Dirac for constrained systems.

Moreover, we have shown that the time-gauge condition on the Lagrangian density (\ref{Lag_new}) renders a non-degenerate symplectic matrix $\mathcal{F}^{\text{(final)}}_{ij}$ (\ref{Ff}), whose determinant allows us to identify the functional measure for determining the quantum transition amplitude according to Toms \cite{Toms}, and whose inverse allows one to identify the quantization-brackets. As a consequence, we have confirmed that Zwei-Dreibein gravity has two physical degrees of freedom per space point, as expected. 
Our work suggests that this symplectic method can be straightforwardly applied in massive- and bi-gravity theories written in first-order formalism in three- and four dimensions, trivializing the issue of identification of the physical constraints and gauge structure of such theories. 

\section{Acknowledgments}
This work has been supported by the National Council of Science and Technology (CONACyT) of M\'exico under 
postdoctoral Grant No. $290692$. 

\end{document}